\documentstyle[12pt]{article}
\begin{document}

%%%%%%%If you do not have the msbm fonts, delete the following 4 lines
\font\mybb=msbm10 at 12pt
\def\bb#1{\hbox{\mybb#1}}
\def\Z {\bb{Z}}
\def\R {\bb{R}}
\def\E {\bb{E}}
\def\bfo{\mbox{\boldmath $\omega$}}
\def\bfn{\mbox{\boldmath $\nabla$}}
\def\tr{{\rm tr}}
%%%%%%%%%%%%
%%%and replace with the following 2 lines (without %)
%\def\Z {Z}
%\def\R {R}
%%%%%%%%%%
 \def\unit{\hbox to 3.3pt{\hskip1.3pt \vrule height 7pt width .4pt \hskip.7pt
\vrule height 7.85pt width .4pt \kern-2.4pt
\hrulefill \kern-3pt
\raise 4pt\hbox{\char'40}}}
\def\II{{\unit}}
\def\cM {{\cal{M}}}
\def\half{{\textstyle {1 \over 2}}}

\pagestyle{empty}
\rightline{UG-7/99}
\rightline{DAMTP-1999-46}
\rightline{hep-th/9904020}
\vspace{2truecm}
\centerline{\bf  Solitons on the Supermembrane}
\vspace{2truecm}
\centerline{\bf E.~Bergshoeff $^1$ and P.K. Townsend $^2$}
\vspace{.5truecm}
\centerline{$^1$Institute for Theoretical Physics}
\centerline{Nijenborgh 4, 9747 AG Groningen}
\centerline{The Netherlands}
\vspace{.5truecm}
\centerline{$^2$DAMTP, Silver Street, University of Cambridge,}
\centerline{Cambridge CB3 9EW}
\centerline{United Kingdom}
\vspace{2truecm}
\centerline{ABSTRACT}
\vspace{.5truecm}
Energy bounds are derived for planar and compactified M2-branes in a
hyper-K\"ahler background. These bounds are saturated, respectively,
by lump and Q-kink solitons, which are shown to preserve half the
worldvolume supersymmetry. The Q-kinks have a dual IIB interpretation 
as strings that migrate between fivebranes. 
 
\vfill\eject
\pagestyle{plain}
 
\section{Introduction}

Supersymmetric sigma models in 2+1 dimensions with a K\"ahler
target space generally admit static soliton-like `lump' solutions with  
energy $E= |T|$, where $T$ is the topological charge $\int \omega $ 
obtained by integrating the K\"ahler 2-form $\omega$ 
over the image in target space of the 2-dimensional space (see e.g.
\cite{ruback}). If the K\"ahler  target space admits a holomorphic
Killing vector field $k$ then one can perform a `Scherk-Schwarz' (SS)
dimensional reduction to arrive at a `massive' supersymmetric sigma model
in 1+1 dimensions with a scalar potential $V \sim k^2$. This theory admits
`Q-kink'  solutions \cite{AT,paptown} with an energy

\begin{equation}
E = \sqrt {Q_0^2 + Q^2}\, ,
\end{equation}
where $Q_0$ is the Noether charge associated with $k$, and 
$Q = \int i_k\omega$, the integral being taken over the image in target
space of the 1-dimensional space. Because $k$ is holomorphic the 1-form 
$i_k\omega $ is closed, so $Q$ is a topological charge. When
$Q_0 \ne 0$ the Q-kink is a {\sl time-dependent} solution
of the sigma-model field equations. When $Q_0 = 0$ it becomes a standard 
static kink solution.

A 2+1 dimensional supersymmetric sigma model
with a K\"ahler target space has an N=2 supersymmetry
and the topological charge $T$ appears as a central charge in the
supersymmetry algebra. This implies the bound $E \ge |T|$, which is
saturated by the sigma-model lumps. Similarly,
1+1 dimensional massive supersymmetric sigma models 
obtained by SS dimensional
reduction actually have (2,2) supersymmetry, and both $Q_0$ and $Q$
appear in the supersymmetry algebra as central charges. This implies the
bound $E \ge \sqrt {Q_0^2 + Q^2}$, which is saturated by the Q-kinks.

If the K\"ahler target space is actually hyper-K\"ahler then
the topological charge $T$ of the 2+1 dimensional model is just
one of a triplet of topological charges

\begin{equation} 
{\bf T} = \int\! \bfo\, , 
\end{equation}
where $\bfo$ is the triplet of K\"ahler 2-forms. The number
of supersymmetries is also doubled to N=4, and the triplet of 
charges $\bf T$ appear as central charges in the 
N=4 supersymmetry algebra. If the
hyper-K\"ahler space admits a tri-holomorphic Killing vector field $k$ then 
SS dimensional reduction along its orbits yields a (4,4) supersymmetric
massive sigma model in 1+1 dimensions, again with $V \sim k^2$.
The topological charge $Q$ is now one of a triplet of topological charges

\begin{equation}
{\bf Q} = \int i_k \bfo\, ,
\end{equation}
and the four charges $(Q_0, {\bf Q})$ appear as central charges in the (4,4)
supersymmetry algebra. This implies the bound

\begin{equation}
E \ge \sqrt{Q_0^2 + {\bf Q\cdot Q}}\, ,
\end{equation}
which is saturated by the (hyper-K\"ahler) Q-kinks.

There is a close analogy here to N=2 and N=4 supersymmetric Yang-Mills (SYM)
theories in 4+1 and 3+1 dimensions \cite{AT,dorey}. The lumps 
of the 2+1 dimensional
sigma model are similar to the instantonic solitons of the 4+1
SYM theory; for example, they have no fixed scale. The Q-kinks of the 1+1
dimensional sigma model are similar to the dyons of 3+1 SYM theory;
for example the sigma model has a vacuum angle and Q-kinks generally have 
fractional $Q_0$-charge, just as SYM dyons generally carry fractional
electric charge for nonzero vacuum angle. The scale introduced by the potential
term in the 1+1 dimensional sigma model is analogous to the scale introduced
by the Higgs mechanism in the SYM case.

The N=2 and N=4 SYM theories have interpretations
in IIB string theory as effective field theories describing the
fluctuations of D-branes around some `vacuum' brane
configuration. The dyon solutions are the field theory realization
of (p,q) strings, or string webs, stretched between the
D-branes. A feature of the brane
interpretation of the SYM theories is that in a limit in which the 
individual branes become widely separated the dyon solutions
must transmute into a solution of the equations governing
the dynamics of a ${\sl single}$ brane. This is an 
Abelian SYM theory, although not of a conventional
type because the brane action involves higher derivative interactions.
These `DBI solitons', were found in \cite{CM,G}; the supersymmetric
solutions are worldvolume `spikes' of infinite total energy per unit length
equal to the tension of a (p,q) string. Solutions with finite
${\sl total}$ energy can be found by considering the DBI action
in an appropriate supergravity background \cite{UK}.

These considerations motivate us to seek an interpretation of 
sigma-model lumps and Q-kinks as solitons on the worldvolume of the
eleven-dimensional supermembrane \cite{supermem}, otherwise known as the
M2-brane. An M2-brane in a vacuum background has supersymmetric, but infinite
energy, vortex  solutions that  can be interpreted as intersections with other
M2-branes \cite{CM,G,GGT}. In a non-vacuum K{\" a}hler background we may have 
the option of wrapping the `other' M2-branes on finite area holomorphic 2-cycles
of the background. These are finite energy solitons that provide the brane
realization of K{\" a}hler sigma-model lumps. We shall concentrate here on the
hyper-K\"ahler case; specifically, we shall consider the supermembrane in a
background for which the 4-form field-strength vanishes and the 11-metric
takes the form

\begin{equation}
\label{11m}
ds^2 = ds^2(\E^{(1,5)} \times S^1)  + ds^2_4\, ,
\end{equation}
where $ds^2_4$ is the Kaluza-Klein (KK) monopole metric

\begin{equation}\label{tfour}
ds^2_4=  V^{-1}\left (d\varphi - {\bf A}\right )^2 
+ V\, ds^2(\E^3) \, .
\end{equation}
The 1-form $\bf A$ on $\E^3$ satisfies
$\bfn  V = \bfn \times {\bf A}$, which implies that 
$V$ is harmonic on $\E^3$. The vector field $\partial/\partial\varphi$ is
Killing and triholomorphic. We take it to be the vector field $k$
of the previous discussion, i.e.

\begin{equation}
\label{KV}
k = \partial/\partial\varphi\, .
\end{equation}
The orbits of $k$ are
Kaluza-Klein (KK) circles which shrink to points at singularities of $V$.
Let $\bf X$ be Cartesian coordinates on $\E^3$ and ${\bf X}_0$ a constant
3-vector from the origin. The simplest choice of $V$ that serves our 
purposes is

\begin{equation}\label{vee}
V = 1 + {1\over |{\bf X}+{\bf X}_0|} + {1\over |{\bf X}-{\bf X}_0|}\, , 
\end{equation}
which describes a two-centre KK-monopole of M-theory. 

Upon reduction on orbits of $k$, the KK-monopole acquires an  
interpretation as two parallel IIA D6-branes separated in $\E^3$  by the
constant vector $2{\bf X}_0$. The two centres of the metric at 
${\bf X}=\pm {\bf X}_0$ can be considered as the poles of a 2-sphere
parametrized  by $\varphi$ and the distance from one D6-brane along the line
joining the two of them. A membrane wrapped on this 2-sphere has a IIA
interpretation  as a string stretched between the two D6-branes \cite{sen}. 
Now consider a D2-brane parallel to the two D6-branes. In general it will not
be colinear in $\E^3$ with the two D6-branes and so will not intersect the
string joining them. However, we may move it until it does intersect. From the
D=11 perspective we then have a pointlike intersection of two M2-branes, one an
infinite planar one and the other one wrapped on a finite area 2-cycle of the
background. The singular intersection point may be desingularized so that we
have a single M2-brane with a non-singular lump soliton on it of some finite size
$L$. From the IIA perspective this corresponds to separating the points at
which the strings from each of the two D6-branes meet the D2-brane. 

In the case of the lump, the vacuum is an infinite planar M2-brane. To find 
a brane interpretation of the hyper-K\"ahler Q-kink we will need to wrap 
this M2-brane on some one-cycle of the background space. This corresponds 
to SS reduction on some Killing vector field with closed orbits. The 
dimensional reduction will preserve all supersymmetries only if this 
Killing vector field is triholomorphic. The Killing vector field $k$ 
of (\ref{KV}) is therefore an obvious candidate, but SS reduction 
on orbits of $k$ does ${\it not}$ yield a potential $V \sim k^2$ 
as one might have expected from our earlier summary of the results of SS
reduction in sigma models.
Rather, it yields a non-vanishing, and non-uniform, IIA string tension.
The non-uniformity of the tension creates an attractive force between the
string and the D6-brane but on reaching the D6-brane core the string 
can simply dissolve into Born-Infeld flux. 
To get the potential term in the dimensionally reduced action one must 
suppose that the 11-metric (\ref{11m})
has another tri-holomorphic Killing vector field
with closed orbits. 
We may take this to be a vector field generating the U(1) isometry
of the $S^1$ factor in this metric. Let us call this vector field $\ell$. 
Dimensional reduction on 
orbits of $k+\ell$ leads to a bound state of the IIA string discussed
above with a D2-brane wrapped on orbits of $\ell$. This bound state is
itself bound to the D6-brane. The effective string action is the desired
brane version of the massive hyper-K\"ahler sigma model, admitting Q-kink
solutions. T-dualizing in the (compact) $\ell$ direction yields a
(1,1) IIB string bound to a D5-brane. As we shall see, the Q-kink 
solution can then be interpreted as a (1,1) string that migrates 
from one D5-brane to another. 

Although lump and Q-kink solutions are known to minimise the energy of the
relevant sigma model it does not immediately follow that they 
minimise the energy 
on the M2-brane because of the nonlinearities of the Dirac membrane 
action. 
By means of the 
brane version of the Bogomol'nyi argument \cite{GGT}, we show
 that the energy of the M2-brane is indeed minimised by these solutions.
We consider the lumps first,
as these are static, and then generalize to the Q-kinks. Both 
configurations are then shown to preserve some fraction of the 
worldvolume supersymmetry. Again, this is known in the 
sigma-model case, but the 
supersymmetry transformations of the supermem\-brane are different.
They can be deduced from a combination of the target space 
supersymmetry and the kappa-symmetry of the supermembrane,
and this leads to a simple condition for a worldvolume field configuration 
to preserve some fraction of supersymmetry \cite{singleton,bbs,BKOP}. For a
vacuum background this condition is easily interpreted as a constraint on
the 32 independent constant Killing spinors of the background, but its
interpretation is less direct in a non-vacuum background in which the Killing
spinors are not constant and span a space of lower dimension. Here we
present a more geometrical derivation of the conditions for
preservation of supersymmetry and we discuss some 
subtleties of the non-vacuum case that have been passed over previously.

\section{Energy bounds}

Our starting point for finding soliton solutions as minimum energy 
configurations of the supermembrane will be its Hamiltonian 
formulation \cite{yosh}. Let
$\xi^i = (t, \sigma^a)$ be the worldvolume coordinates, with $\sigma^a$
the worldspace coordinates, and let $X^m$ be the D=11 spacetime coordinates.
The supermembrane Lagrangian, omitting fermions, can then be written as

\begin{equation}
{\cal L}  = P_m \dot {X}^m - s^a P_m\partial_a X^m -
{\textstyle{1\over 2}} v \left [ P^2 + {\rm det} (g_{ab}) \right ]\, ,
\end{equation}
where

\begin{equation}
g_{ab} = \partial_a X^m \partial _b X^n g_{mn}
\end{equation}
is the induced worldspace metric, $P_m$ is the 11-momentum
conjugate to $X^m$, and $s^a$ and $v$ are Lagrange multipliers. Let
$X^m = (Y^i, X^I)$ ($i=0,1,2$) so that

\begin{equation}
ds^2_{11} = dY^i dY^j \eta_{ij} + dX^I dX^J g_{IJ}\, ,
\end{equation}
where $\eta$ is the 3-dimensional Minkowski metric. We make the 
gauge choice $Y^i(\xi) = \xi^i$. This implies that

\begin{equation}
g_{ab} = \eta_{ab} + \partial_a X^I \partial_b X^J g_{IJ}\, .
\end{equation}
It also implies that

\begin{equation}
P_m = \left ( -\varepsilon -1, -P_I\partial_a X^I, P_I\right )
\end{equation}
where $\varepsilon$ is the energy density relative to the brane vacuum 
(which is taken to have unit tension). The Hamiltonian constraint 
imposed by $v$ can be solved for $\varepsilon$

\begin{eqnarray}\label{16}
(\varepsilon +1)^2 &=& 1 + \nabla X^I \cdot \nabla X^J g_{IJ} + (g^{IJ} +
\nabla X^I \cdot \nabla X^J)P_I P_J\nonumber\\
&&+ {\textstyle{1\over 2}} (\nabla X^I \times \nabla X^J)(\nabla X^K\times
\nabla X^L)g_{IK}g_{JL}\, ,
\end{eqnarray}
where we have used standard 2D vector calculus notation for worldspace 
derivatives. This expression differs in several respects from the 
corresponding expression for the sigma-model energy density. Firstly, the 
supermembrane expression is quadratic in $\varepsilon$; this is because 
the sigma-model approximation is a kind of non-relativistic approximation to 
the supermembrane (they differ in the same way that the energies of 
a relativistic and non-relativistic particles differ). Secondly, 
the supermembrane expression involves terms quartic in derivatives that are
absent in the sigma-model case.  

\subsection{Lumps}

We now aim to rewrite the above expression for the energy density in the form

\begin{eqnarray}
\label{e2}
(\varepsilon+1)^2 &=& \left ( 1 \pm {\textstyle {1\over 2}}\nabla X^I \times 
\nabla X^J \omega_{IJ}\right )^2 \nonumber\\
&&+ {\textstyle {1\over 2}}\left ( \nabla X^I \pm *\nabla X^K I_K{}^I\right)
\left ( \nabla X^J \pm *\nabla X^L I_L{}^J\right ) g_{IJ}\\
&&+ {\textstyle{1\over 4}}\sum_{r=1}^6 \left 
(\nabla X^I \times \nabla X^J \Omega_{IJ}^{(r)}\right )^2\, ,
\nonumber
\end{eqnarray}
where we have set $P_I =0$ and
$*\nabla = (\partial_2, -\partial_1)$ if $\nabla = 
(\partial_1, \partial_2)$. We assume that $I_I{}^J$ is a complex
structure, that the 8-metric $g_{IJ}$ is Hermitian with respect to it
and that $\omega_{IJ}=I_I{}^Kg_{KJ}$ is the corresponding closed 
K\"ahler 2-form. For the moment we leave unspecified the six 
2-forms $\Omega^{(r)}$. These conditions are already  sufficient to 
ensure that 
all but the quartic terms in $\nabla X$ of (\ref{16}) are reproduced. 
To reproduce the quartic terms too we require that

\begin{equation}
\label{XIJ}
X^{IJ} X^{KL}
\left [ \omega_{IJ} \omega_{KL} + \sum_{r=1}^6 \Omega^{(r)}_{IJ} 
\Omega^{(r)}_{KL}
 -2 g_{KI}g_{JL}\right ] = 0\, ,
\end{equation}
where
\begin{equation}
X^{IJ} \equiv \nabla X^I \times \nabla X^J\, .
\end{equation}
Note that $X^{IJ}$ is an antisymmetric $8\times 8$ matrix. If none
of its 4 skew-eigenvalues vanish, then (\ref{XIJ}) implies that

\begin{equation}
\label{15}
\omega_{I(J} \omega_{K)L} + \sum_{r=1}^6 
\Omega^{(r)}_{I(J} \Omega^{(r)}_{K)L} = g_{I(K}g_{J)L}
- g_{KJ}g_{IL}\, .
\end{equation}
For a membrane in flat space this condition is satisfied by taking the
matrices
\begin{equation}
I_I{}^J \equiv \omega_{IK}g^{KJ}\, ,\qquad 
(J^{(r)})_I{}^J \equiv \Omega^{(r)}_{IK}g^{KJ}
\end{equation}
to be the seven complex structures of $\E^8$. 

For every vanishing skew eigenvalue of $X^{IJ}$ the dimension of the
transverse space is effectively reduced by two. In this reduced space, 
we must again have (\ref{15}) but it may now be possible to choose some 
of the six $J$ matrices to vanish. For example, if $X^{IJ}$ has two vanishing
skew-eigenvalues then the transverse space is effectively 4-dimensional; in
other words, there are four `active scalars'. We may now set all but two of
the $J$ matrices to zero. The other two, together with $I$ can be taken to
be the three almost complex structures of the transverse 4-manifold (these
will be covariantly constant if this transverse 4-space is hyper-K\"ahler,
but we need not assume any special properties at this point).
If $X^{IJ}$ has three vanishing skew-eigenvalues, corresponding to two
active scalars, then the transverse space is effectively two-dimensional, 
and we may take all the $J$ matrices to vanish. 
 
Given (\ref{XIJ}) we deduce that

\begin{equation}
\varepsilon  \ge {\textstyle{1\over 2}}|X^{IJ}  \omega_{IJ}|
\end{equation}
with equality when

\begin{equation}\label{17}
\nabla X^I = \mp *\nabla X^J I_J{}^I
\end{equation}
and

\begin{equation}
\label{18}
X^{IJ} \Omega_{IJ}^{(r)} = 0\hskip 1truecm r = 1,\cdots ,6\, .
\end{equation}

The condition (\ref{17}) is the statement that in complex coordinates
$Z^\alpha$, adapted to the complex structure $I$, the functions 
$Z^\alpha(z)$ are holomorphic on worldspace, with 
$z=\sigma^1 + i\sigma^2$. The conditions (\ref{18}) are implied by 
(\ref{17}) if the matrices $J^{(r)}$ are such that

\begin{equation}
\label{19}
IJ^{(r)} + J^{(r)} I = 0\, ,\hskip 1.5truecm r = 1,\cdots ,6\, .
\end{equation}
This is true when $I,J^{(r)}$ are the 7 complex structures
of $\E^8$. It is also satisfied if $I,J^{(1)}, J^{(2)}$ are the three
almost complex structures of a 4-dimensional space, with the other $J$
matrices vanishing. This is the case of most interest here because
we may obviously reduce the transverse 8-space to an effective transverse
4-space by requiring all scalars to vanish except those
associated with the $ds^2_4$ metric in (\ref{11m}). This restriction still
allows configurations with either two or four active scalars. 

In the case of a flat background, a solution of (\ref{17}) with $2n$ real
`active scalars' has the interpretation as the (orthogonal) intersection with
the  worldvolume of $n$ M2-branes, corresponding to a spacetime intersection of
$n+1$ M2-branes. The spacetime configuration is known to preserve the
fraction $1/2^{n+1}$ of the spacetime supersymmetry \cite{mbrane} so we may 
expect the fraction of worldvolume supersymmetry preserved to be $1/2^n$. This
can be confirmed directly from a consideration of $\kappa$-symmetry of the
supermembrane \cite{gibpap,GLW}. The lump solution of (\ref{17}) for
the KK-monopole background is also one with two `active scalars' and preserves
half the worldvolume supersymmetry but the total number of worldvolume
supersymmetries is half what it would be in a flat spacetime. The fraction of
supersymmetry of the M-theory vacuum that is preserved by the total system is
therefore 1/8 (1/2 for the solution, 1/2 for the brane and 1/2 for the
background). We shall examine the question of supersymmetry in more
detail in section 3.

\subsection{Q-Kinks}

We now set

\begin{equation}\label{ssred}
\partial_2 X^I = k^I\, ,
\end{equation}
where $k$ is a holomorphic Killing vector field. The holomorphicity
condition ensures that the dimensionally
reduced 1+1 dimensional theory preserves the $N=2$ supersymmetry
of the (2+1)-dimensional model. Any additional supersymmetries will 
be associated with additional complex structures; if $k$ is holomorphic 
with respect to them too then the reduction will preserve these 
additional supersymmetries. For the KK-monopole background we may take
$k$ to be the triholomorphic Killing vector of (\ref{KV}). 
Using (\ref{ssred}) in (\ref{16}) we have

\begin{eqnarray}\label{24}
(\varepsilon+1)^2 &=& 1 + \left ( g^{IJ} + 
\partial X^I\partial X^J + k^I k^J\right )
P_I P_J + |\partial X|^2 + |k|^2\nonumber\\
&& + \ 2 \partial X^{[I} k^{J]} \partial X^{[K} k^{L]} g_{IK} g_{JL}\, ,
\end{eqnarray}
where $\partial X = \partial_1 X$. Restricting to static $(P=0)$ and
uniform $(\partial X = 0)$ configurations yields
$\epsilon = \sqrt{1 + |k|^2} - 1 \approx {1\over 2} |k|^2$,
which is the membrane version of the scalar potential that leads to
Q-kink solutions interpolating between its
minima at fixed points of $k$ where $|k|$ vanishes.

Under the same conditions as before, 
the expression (\ref{24}) for the energy density can be rewritten as

\begin{eqnarray}
(\varepsilon+1)^2 &=& \left [ 1 + v k\cdot P + 
\sqrt {1-v^2}\ \partial X^I  k^J
\omega_{IJ}\right ]^2 + \left |P - vk\right |^2\nonumber\cr
&&+\ \left |\partial X^I + \sqrt {1-v^2}\ k^J I_J{}^I\right |^2 +
\left (P\cdot \partial X\right )^2\cr
&&+\ \left [v\ \partial X^I k^J \omega_{IJ} -
\sqrt {1-v^2}\ k\cdot P\right ]^2
+ \sum_r \left ( \partial X^I k^J \Omega_{IJ}^{(r)}\right )^2\, .
\end{eqnarray}
for arbitrary constant $v$ with $|v|<1$. 
We deduce that

\begin{equation}
\varepsilon 
\ge vk\cdot P + \sqrt {1-v^2}\ \partial X^I k^J \omega_{IJ}\, 
\end{equation}
with equality when

\begin{eqnarray}\label{kinkbps}
P^I &=& v k^I\, ,\cr
\partial X^I &=& - \sqrt {1- v^2}\ k^J I_J{}^I\, ,
\end{eqnarray}
since these equations imply the vanishing of the remaining terms.

Setting $P^I=\dot X^I$ in (\ref{kinkbps}) we recover the equations found in 
\cite{AT}, the solutions of which are Q-kinks. The explicit Q-kink 
solution of \cite{AT} was given for the two-centre metric with $V$ as in
(\ref{vee}) but without the constant term (i.e. for the Eguchi-Hanson 
metric \cite{prasad}). The explicit
solution when $V$ includes a constant term has been found by 
Opfermann \cite{andreas}.

\section{Supersymmetry}

The supermembrane is invariant under all isometries of the background.
Supersymmetries correspond to Grassmann odd Killing vector superfields
$\chi = \chi^A E_A$, where $E_A = E_A{}^M\partial_M$. The (Grassman even) 
spinor component $\chi^\alpha$ is a Killing spinor in the standard sense, 
at least 
in a purely bosonic background. The (Grassman odd) vector 
component $\chi^a$ is a 
superfield satisfying the constraint ${\cal D}_\alpha \chi^a = 
(\Gamma^a\chi)_\alpha$.  Let
$\{\chi\}$ be the complete set of these Killing vector superfields and let
$\{\epsilon\}$ be a corresponding set of anticommuting parameters.
The supersymmetry transformations of the worldvolume fields $Z^M$ are then

\begin{equation}
\delta_\epsilon Z^M = \epsilon\cdot \chi^M\, ,
\end{equation}
where $\epsilon\cdot\chi$ is used to denote the sum over the $(\epsilon,\chi)$
pairs. Defining $\delta E^A = \delta Z^M E_M{}^A$, we then have

\begin{equation}
\delta_\epsilon E^A = \epsilon\cdot \chi^A\, .
\end{equation}
The $\kappa$-symmetry variation $\delta_\kappa Z^M$ can be similarly 
expressed in the form

\begin{equation}
\delta_\kappa E^\alpha = \kappa^\beta (1+\Gamma)_\beta{}^\alpha\, ,
\hskip 1truecm \delta_\kappa E^a = 0\, ,
\end{equation}
where
\begin{equation}
\Gamma = {1\over 6\sqrt{-g}}\epsilon^{ijk} E_i{}^a E_j{}^b E_k{}^c
\Gamma_{abc}\, ,
\end{equation}
with $E_i{}^a = \partial_i Z^M E_M{}^a$, and $g$ is the determinant of 
the induced worldvolume metric $g_{ij} = E_i{}^a E_j{}^b\eta_{ab}$. To
fix $\kappa$-symmetry, we make the gauge choice \cite{singleton}

\begin{equation}
E^\alpha (1+\Gamma)_\alpha{}^\beta =  0\, .
\end{equation}
This restricts only $dZ^M$, but this is sufficient. 
Note that this gauge choice is invariant under supersymmetry, at 
least in a bosonic background and for vanishing worldvolume fermions; under 
these conditions we may neglect the variation of $\Gamma$, while

\begin{equation}
\delta_\epsilon (dZ^M E_M{}^\alpha) = D(\epsilon\cdot\chi)^\alpha - 
\epsilon\cdot\chi^\beta E^\gamma T_{\gamma\beta}{}^\alpha \, ,
\end{equation}
which vanishes by the Killing spinor equation (the $T_{\gamma\beta}{}^\alpha$
component of the torsion tensor is proportional to the 4-form field strength of
D=11 supergravity).

The remaining physical variables are such that their
variations are $\delta E^\alpha (1-\Gamma)_\alpha{}^\beta$.
The condition that the worldvolume configuration preserves some
supersymmetry is therefore 
\begin{equation}
\label{30}
\epsilon \cdot \chi^\alpha (1-\Gamma)_\alpha{}^\beta = 0\, .
\end{equation}
For flat superspace, $\chi_I{}^\alpha = \delta_I{}^\alpha$, so
$\epsilon\cdot\chi=\epsilon$, a constant 32-component spinor.
We thus recover the flat space condition \cite{singleton}
\begin{equation}\label{flat}
\epsilon^\alpha (1-\Gamma)_\alpha{}^\beta = 0\, .
\end{equation}
More generally, we must take into account the fact that  $\epsilon\cdot\chi$ is
neither constant nor a spinor with 32 independent components. For the simplest
backgrounds, including the KK-monopole background considered here, we have
\begin{equation}\label{first}
\epsilon\cdot \chi = f_\chi\, \epsilon\, ,
\end{equation}
where $f_\chi$ is an ordinary function, and $\epsilon$ is a constant 
32-component spinor satisfying
\begin{equation}\label{chicon}
P_\chi\, \epsilon =0\, ,
\end{equation}
with $P_\chi$ a {\sl constant} projection matrix. For the 
KK-monopole background
the matrix $P_\chi$ is just the product of four constant 
Dirac matrices, one for
each of the four dimensions of the 4-metric, and it 
has the property (associated
with the fact that this background preserves 1/2 of the spacetime
supersymmetry) that $\tr P_\chi=16$. The fraction of spacetime supersymmetry
preserved by the brane plus background configuration is therefore determined by
the number of simultaneous solutions to (\ref{flat}) and (\ref{chicon}).
Note that the function $f_\chi$ of (\ref{first}) is 
irrelevant to the final result.

We now fix worldvolume diffeomorphisms by the `static gauge' choice

\begin{equation}
X^m = \left (\xi^i, X^I(\xi)\right )\, .
\end{equation}
With this gauge choice the condition (\ref{flat}) becomes

\begin{eqnarray}\label{gaugefixed}
\sqrt {-g}\, \epsilon &=& \big[\Gamma_* + 
\Gamma^k\partial_k X^I\Gamma_I \Gamma_*
+ {\textstyle{1\over 2}}\Gamma_k \epsilon^{ijk}\partial_iX^I\partial_jX^J
\Gamma_{IJ}  \nonumber\\
&& +\ {\textstyle{1\over 6}}\varepsilon^{ijk}\partial_i 
X^I\partial_j X^J \partial_k X^K \Gamma_{IJK}\big]\,\epsilon\, ,
\end{eqnarray}
where
\begin{equation}
\Gamma_* = \Gamma_{012}\, .
\end{equation}
In addition 

\begin{equation}
g = {\rm det}\left (\eta_{ij} + {\tilde g}_{ij}\right )\, ,
\end{equation}
where

\begin{equation}
{\tilde g}_{ij} = \partial_i X^I\partial_j X^J g_{IJ}\, .
\end{equation}

The condition (\ref{gaugefixed}) for preservation of 
supersymmetry can now be expanded in a power series in 
$\partial X$. We assume here that each term in the series must vanish
separately\footnote{For a flat space background this amounts to the
assumption that the worldspace is the contact set of a K{\"a}hler 
calibration. K{\"a}hler calibrations are only ones of relevance here, although 
for the M5-brane there are other calibrations for which the
assumption would be false. See \cite{gibpap,GLW,QMW} for a discussion
of calibrations in relation to branes.}.  
At zeroth order in this expansion we learn that

\begin{equation}
\Gamma_* \epsilon = \epsilon\, .
\end{equation}
Because the projector $P_\chi$ involves only the $\Gamma^I$ matrices, this
equation tells us that the worldvolume vacuum preserves 
half the supersymmetries
of the supergravity background, i.e. that the M2--brane 
is $1/2$ supersymmetric.

At first order in the $\partial X$ expansion we learn that

\begin{equation}\label{key}
\Gamma^k \partial_k X^I \Gamma_I\, \epsilon = 0\, .
\end{equation}
This implies various higher-order identities. In particular
it implies that ${\rm det}\, {\tilde g}_{ij}$ vanishes and that

\begin{equation}
\label{38}
\eta^{im}
\eta^{jn} {\tilde g}_{jm} {\tilde g}_{in}
 = {\textstyle{1\over 2}}(\eta^{ij}{\tilde g}_{ij})^2\, .
\end{equation}
Using these identities, and the constraints on $\epsilon$ quadratic 
and cubic in
$\partial X$ that also follow from (\ref{key}), one can show that the 
${\sl full}$ constraint $\Gamma\epsilon = \epsilon$ is satisfied.
Thus, (\ref{key}) is the only condition (apart from $P_\chi\epsilon=0$) that
we need analyse to determine the fraction of supersymmetry preserved by  
lump and Q-kink soliton solutions. 

Having found the conditions for partial preservation of worldvolume
supersymmetry, we are now in a position to verify that the lump and Q-kink
solitons are supersymmetric and to determine the fraction of supersymmetry they
preserve. We need not discuss the lump and Q-kink cases separately
because the
formalism to follow will apply equally to both. For lumps we just set $v=0$ 
while for Q-kinks we set $\partial_2 X^I = k^I$. We begin with the observation
that the equations

\begin{equation}
\label{39}
\dot {X}^I = v\partial_2 X^I\, ,\hskip 1.0truecm
\partial_1 X^I + \sqrt {1-v^2}\ \partial_2X^J\, I_J{}^I=0\, ,
\end{equation}
imply that

\begin{equation}
{\tilde g} = \tilde g_{22} \times \pmatrix{v^2&0&v\cr
                      0&1-v^2&0\cr
                      v&0&1}\, ,
\end{equation}
which manifestly has vanishing determinant and solves (\ref{38}).
Using (\ref{39}) in (\ref{key}) and 
$\Gamma_*\epsilon = \epsilon$,
we have

\begin{equation}\label{susycon}
\partial_2 X^I\Gamma^J
\left ( g_{IJ} + {\tilde \Gamma} \omega_{IJ}\right )\epsilon = 0\,,
\end{equation}
where

\begin{equation}
{\tilde\Gamma} = {1\over \sqrt{1-v^2}}\left (\Gamma^0 + v \Gamma^2
\right )\, .
\end{equation}
Note that $\tilde \Gamma^2=-1$. 

The matrices $\Gamma^I$ are space-dependent. We can write them 
in terms of the constant complex matrices 
\begin{equation}
\Gamma^\alpha = \Gamma^I e_I{}^\alpha 
\end{equation}
and their complex conjugates $\bar\Gamma^{\bar\alpha}$, where $e_I{}^\alpha$
is a complex target space vielbein (with complex conjugate
$\bar e_I{}^{\bar\alpha}$) chosen such that
\begin{equation}
\{\Gamma^\alpha,\Gamma^\beta\}=0 \qquad 
\{\Gamma^\alpha,\bar\Gamma^{\bar\beta}\} = \delta^{\alpha\bar\beta}\, .
\end{equation}
Using the fact that $\omega_{\alpha\bar\beta}= 
i\delta_{\alpha\bar\beta}$ in this
basis, we now have
\begin{equation}
 \sum_{\alpha=\bar\alpha} \left[ e^\alpha \bar\Gamma^{\bar\alpha} + 
\bar e^{\bar\alpha} \Gamma^\alpha + i(e^\alpha\bar\Gamma^{\bar\alpha} - 
\bar e^{\bar\alpha}\Gamma^\alpha)\tilde \Gamma \right ]\epsilon =0
\end{equation}
where
\begin{equation}
e^\alpha = \partial_2 X^I e_I{}^\alpha\, .
\end{equation}
Each term in the sum must vanish separately. This leads to a set of equations,
each of which can be written in the form
\begin{equation}
(e\bar\Gamma + \bar e\Gamma)(1-i\tilde 
\Gamma[\Gamma,\bar\Gamma])\epsilon =0\, .
\end{equation}
It follows that {\it either} $e=0$, which effectively requires one 
complex worldvolume scalar to be constant, {\it or} $\epsilon$ must 
satisfy the constraint
\begin{equation}
(1-i\tilde \Gamma[\Gamma,\bar\Gamma])\epsilon =0\, ,
\end{equation}
which reduces the fraction of supersymmetry preserved by two, unless it is
already satisfied by virtue of the $P_\chi$ projection imposed by the
background. 

We briefly discussed the flat background case in section 2.1. Solutions with 
$2n$ active (real) scalars preserve $1/2^n$ of the 
worldvolume supersymmetry and
hence $1/2^{n+1}$ of the spacetime supersymmetry; 
their spacetime interpretation
is as $n+1$ intersecting M2-branes.  The computation of the fraction of
worldvolume supersymmetry preserved by the finite energy lumps and Q-kinks is
slightly more involved because the effects of the $P_\chi$ projection must be
taken into account. However this just reduces the initial number of
supersymmetries by a factor of two. The M2-brane breaks half of that and 
the lump and Q-kink solitons halve it again, 
exactly as in the flat space case. 

These results could be anticipated from the central charge structure of the
supermembrane worldvolume superalgebra. In the KK-monopole background we would
need to consider the N=4 D=3 worldvolume supersymmetry algebra. For simplicity
we concentrate here on the N=8 D=3 algebra relevant to a 
supermembrane in a flat
space background. As we are considering only scalar central charges, the
supersymmetry algebra is \cite{BGT}
\begin{equation}
\{Q_\alpha^{\tilde I} ,Q_\beta^{\tilde J}\} = 
\delta^{\tilde I \tilde J}P_{\alpha\beta} + 
\varepsilon_{\alpha\beta}\tilde Z^{\tilde I \tilde J}\, ,
\end{equation}
where the 8 supersymmetry charges $Q^{\tilde I}$ transform as a chiral $SO(8)$
spinor. The antisymmetric central charge matrix $\tilde Z$ has four skew 
eigenvalues $\zeta_k$ ($k=1,2,3,4$). The positivity
of the $\{Q,Q\}$ anticommutator implies the bound
\begin{equation}
M^2 \ge {\rm sup} (\zeta_1,\zeta_2,\zeta_3,\zeta_4)\, .
\end{equation}
The fraction of worldvolume supersymmetry preserved is $2^{n-5}$ where $n$ is
the number of factors of $\det \{Q,Q\}$ of the form $(M^2-\zeta)^2$ that
simultaneously vanish. For example, states for which all four 
skew eigenvalues are equal, but non-zero, preserve half of the 
worldvolume supersymmetry. Note that since $\tilde I$ is a {\sl spinor}
index the central charge $\tilde Z$ cannot be directly interpreted as the
two-form topological charge $Z$ associated with a membrane in a given
2-plane; the relation between the two is such that equal skew-eigenvalues
of $\tilde Z$ corresponds to three vanishing skew-eigenvalues of $Z$, and
vice-versa.

\section{IIB interpretation of Q-kinks}

In the introduction we explained briefly the IIA superstring 
interpretation of the supermembrane lump solutions. As mentioned there,
the most natural superstring interpretation of Q-kinks is in terms of IIB
superstring theory. We now return to this point.

It was implicit in our discussion of the Q-kink in section 2.2 
that $\xi^2=\rho$ is periodically identified; otherwise we do not have 
a genuine compactification.
Since we had already made the static gauge choice $Y^2= \rho$, it follows
that we must take $Y^2$ to be an angular variable, i.e. the coordinate of
the $S^1$ factor in (\ref{11m}). In fact, the Killing
vector field $\partial/\partial Y^2$ can be identified as a multiple
of the triholomorphic Killing vector field $\ell$ mentioned in the 
introduction. A standard dimensional reduction on orbits of the 
triholomorphic Killing vector field $k$ would imply that $Y^2$ is 
the only field depending on $\rho$. However, the SS reduction 
ansatz of (\ref{ssred}) means that $Y^2$ is not the only $\rho$-dependent 
worldvolume field. In fact, given (\ref{KV}), the condition
(\ref{ssred}) implies that $\partial_\rho\varphi=1$, or $\varphi=\rho$ up 
to a constant. If we introduce the new coordinates
\begin{equation}
X^0 = \varphi - Y^2 \qquad \tilde Y = {1\over2}\left(Y^2 + \varphi\right)
\end{equation}
then the combination of the static gauge choice and the SS reduction imply that
$(X^0,{\bf X})$ are $\rho$-independent while $\tilde Y=\rho$. In other words,
we are wrapping the membrane on the $\tilde Y$ direction, i.e. on the 
$k+\ell$ cycle. We can consider this to be a non-marginal bound state of a 
membrane wrapped on the $k$ cycle with one wrapped on the $\ell$ cycle
\cite{george}. The IIA interpretation of this bound state (with the $k$
cycle interpreted as the KK circle) was explained briefly in the introduction:
a membrane wrapped on the $k$ cycle yields a IIA 
string in the D6-brane while a membrane wrapped on the $\ell$ cycle yields
a D2-brane, so we end up with a IIA string bound to a D2-brane in a D6-brane.
We now consider the IIB interpretation obtained by T-duality in the $\ell$
direction.

The IIB dual of a membrane wrapped once on each of the two cycles of the 
torus relating the IIB theory to M-theory is a (1,1) string \cite{jhs}.
In our case the (1,1) string is bound to the D5-brane that is
the IIB dual of the KK-monopole. The binding is due to the
fact that the D5-brane attracts (1,0) strings and 
is neutral to (0,1) strings. Thus, there is 
effectively a potential confining the (1,1) string to the D5-brane
(as expected from the $V\sim k^2$ potential relevant to the IIA
description; in fact, the potential is T-duality 
invariant \cite{lambert}). Given sufficient 
energy, the (1,1) string could migrate from one D5-brane to another one
at some position in the transverse 4-space specified by a 4-vector. In fact
the supermembrane Q-kinks discussed earlier correspond to strings which begin
on one D5-brane but then jump over to another one. The charge 4-vector
$(Q_0,{\bf Q})$ is just the position 4-vector of the other D5-brane, as we 
now explain. 

The triplet of K\"ahler 2-forms associated with the 4-metric (\ref{tfour}) is
\begin{equation}
{\bfo} = (d\varphi + {\bf A} \cdot d{\bf X}) d{\bf X} - 
V d{\bf X} \times d{\bf X}
\end{equation}
where the wedge product of forms is understood. Hence the triplet of 
topological charges ${\bf Q}$ is given by
\begin{equation}
{\bf Q} = \int i_k\bfo = \int d{\bf X} 
\end{equation}
where the integral is over the (1,1) string worldspace. For $V$ as given in 
(\ref{vee}), the potential $k^2$ has minima at 
${\bf X}=\pm {\bf X}_0$, so a string that
starts at one minimum and ends at the other one has a 3-vector kink charge
${\bf Q} = 2{\bf X}_0$.
This is the same charge as in the IIA interpretation. However, the 
Noether charge in the IIA interpretation becomes
a fourth topological charge in the IIB interpretation
(cf. \cite{lambert}). To see this it is 
simplest to get to the IIB theory by first compactifying on the $\ell$ cycle 
followed by T-duality on the $k$ cycle. This leads to the S-dual of the 
configuration obtained 
from performing these operations in the reverse order
(i.e. a (1,1) string in a NS-5-brane),
but the result we are aiming at is unaffected by S-duality. Having compactified
on the $\ell$ cycle, T-duality on the $k$ cycle takes $\dot\varphi$ to 
$\partial \tilde\varphi$, where $\tilde\varphi$ is the T-dual coordinate, 
and hence takes the Noether charge $Q_0 = \int V^{-1}
\dot\varphi$ to the topological
charge 
\begin{equation}
\tilde Q_0 = \int d\tilde\varphi\, .
\end{equation}
This result is to be expected
from the fact that the transverse space of the IIB D5-brane is 4-dimensional.
Thus, in the IIB theory the Q-kink charges $(Q_0,{\bf Q})$ become a single
topological 4-vector charge $Q=(\tilde Q_0,{\bf Q})$. A configuration for 
which this charge is non-zero represents a (1,1) string that starts at
one D5-brane and then migrates to another one positioned at some distance
$|Q|$ from the first in the direction given by $Q$.

\vskip .5truecm

\noindent {\bf Acknowledgements}

\vspace{.5truecm}

One of us (PKT) would like to thank Groningen university for its
hospitality, and Neil Lambert and George Papadopoulos for discussions.
The work of E.B.~is supported by the European Commission TMR program
ERBFMRX-CT96-0045, in which E.B.~is associated to the University of
Utrecht.

\end{document}